# Contactless Probing of the Intrinsic Carrier Transport in Single-Walled Carbon Nanotubes

**Yize Stephanie Li[#], Jun Ge, Jinhua Cai, Jie Zhang[&], Wei Lu, Jia Liu, and Liwei Chen[*]**

Suzhou Institute of Nano-Tech and Nano-Bionics, Chinese Academy of Sciences, Suzhou, Jiangsu 215123, P. R. China

[#] Present Address: Department of Materials Science and Engineering, University of Wisconsin-Madison, Madison, WI 53706, USA

[&] Present Address: Molecular Foundry, Lawrence Berkeley National Laboratory, Berkeley, CA 94720, USA

[*] Address correspondence to lwchen2008@sinano.ac.cn

**Abstract:** Intrinsic carrier transport properties of single-walled carbon nanotubes are probed by two parallel methods on the same individual tubes: the contactless dielectric force microscopy (DFM) technique and the conventional field-effect transistor (FET) method. The dielectric responses of SWNTs are strongly correlated with electronic transport of the corresponding FETs. The DC bias voltage in DFM plays a role analogous to the gate voltage in FET. A microscopic model based on the general continuity equation and numerical simulation is built to reveal the link between intrinsic properties such as carrier concentration and mobility and the macroscopic observable, i.e. dielectric responses, in DFM experiments. Local transport barriers in nanotubes, which influence the device transport behaviors, are also detected with nanometer scale resolution.

Investigations on the electronic property of nanomaterials and their potential applications in nanoelectronic, optoelectronic and bioelectronic devices have been major driving forces behind the explosive development of nanotechnology in the last two decades [1–6]. Since the demonstration of the first single-walled carbon nanotube (SWNT) field effect transistor (FET) [7, 8], it has been a standard practice to study the electronic properties of nanomaterials by making metal contacts and extracting carrier concentration and mobility from FET device transport measurements. Significant successes have been achieved, including identification of metallic or semiconducting nature, determination of the electronic scattering mechanism, and unveiling of quantum mechanical phenomena [1, 2]. This approach, however, has several drawbacks such as the challenge of distinguishing the influence of contacts from the property of the nanomaterials themselves, and the difficulty of measuring an ensemble of individual nanomaterials for practical applications involving a large quantity of nanomaterials.

Here we demonstrate that a dielectric response-based approach is capable of studying the electronic property of nanomaterials without making metal contacts. The origin of the dielectric response of any materials to external electric fields may be classified as the classical dipolar response and the carrier dominated low-frequency response [9]. In semiconductors and conductors, the dielectric response is dominated by the contribution from mobile carriers. Thus dielectric property of semiconductors reflects the electronic property of the material, and dielectrics measurement has been an important tool for understanding carrier characteristics in semiconductor research and industry [9–12]. The challenge for characterizing the electronic property of nanomaterials using the dielectric approach lies in the detection of the small dielectric polarization from materials in nanometer dimensions because the dielectric response is basically an extensive quantity proportional to the size or amount of the material. We realize the measurement of small dielectric response of nanomaterials using a scanning force microscopy (SFM) -based force detection, which we have dubbed as dielectric force microscopy (DFM) [13]. The force sensitivity of SFM has reached down to 1 pN, which roughly equals to the force between a single unitary charge and an electric dipole moment of 5 Debye separated by 3 nm. This high sensitivity has enabled the detection of dielectric polarization of small nanometer scaled materials such as individual SWNTs and ZnO nanowires [13–15]. Furthermore, the DFM technique inherits the nanometer-scaled spatial resolution of SFM, and thus allows for spatial mapping of the electronic property of nanomaterials.

In this article, we compare the DFM response and FET device transport measurements on the same SWNTs and establish a quantitative correlation. The intrinsic parallel between these two approaches is explicated through a microscopic Drude-level model. The advantage of the DFM approach on spatial mapping is exemplified in the identification of the location and behavior of local defects in SWNTs.

SWNTs used in this experiment included spin-casted laser ablation nanotubes and horizontally-aligned nanotubes grown by chemical vapor deposition (CVD) [16, 17], both on degenerately p-doped Si wafers with a 300 nm or 1 μm thermal oxide layer. In some experiments, SWNTs were first located with tapping mode atomic force microscopy (AFM) and DFM experiments were carried out on individual clean tubes in ambient air (relative humidity 10% ~ 15%) before device fabrication and transport measurement. In other experiments, FET devices were fabricated and transport properties were measured first, and then the tubes were cut from the electrodes with an AFM probe and DFM measurements were conducted. SWNT-FET devices with a typical back gate geometry, as shown in Fig. 1(a), were fabricated using standard electron beam lithography or photolithography (for as grown CVD tubes). The source and drain electrodes with 5 nm Cr followed by 200 nm Au were deposited by electron beam evaporation. The device was then annealed at 200 ˚C for 10 minutes under the protection of $N_2$ gas to improve the contacts. Electronic transport measurement was carried out in air using a Semiconductor Characterization System (Keithley 4200).

A Park XE-120 AFM was used in imaging experiments, and conducting AFM tips (NSC18/Ti-Pt and NSC19/Ti-Pt, Mickomasch) with a resonance frequency of 75 ~ 80 kHz and spring constant of 1.5 ~ 5.0 N/m, as calibrated experimentally, were used as the probes. A double pass imaging operation is employed [13], as illustrated schematically in Fig. 1(b). A bias voltage $V = V_{DC} + V_{AC}\sin(\omega t)$ from an external function generator (Agilent 33522A) was applied on the conductive AFM tip only in the second pass. Typically, $V_{AC} = 6$ V and $V_{DC}$ was varied between -3 V and 3 V, and $\omega$ was set to be 1 kHz. The AC bias polarizes the sample and results in an attractive dielectric force oscillating in $2\omega$ frequency [13]. The DC bias essentially serves as a local gate that modulates the carrier density in the sample. The $2\omega$ component of the cantilever deflection signal was sampled by a lock-in amplifier (Stanford SR830) and recorded as the dielectric image. The dielectric force was then obtained through probe calibration. Our previous experiment illustrates that DFM signal of metallic SWNTs is essentially independent of $V_{DC}$ while DFM signal of semiconducting SWNTs (ZnO nanowires) increases (decreases) as $V_{DC}$ is decreased. This is interpreted as an indication that semiconducting SWNTs (ZnO nanowires) are p-type (n-type) semiconductors, which are consistent with their carrier types determined from conventional transport measurement [1, 2, 18]. However, a direct proof of the consistency between the DFM and transport measurements on the same nanotubes is still lacking.

Figures 2 and 3 present DFM and FET transport measurements on the same metallic and semiconducting SWNTs, respectively. The DFM signals of the nanotube shown in Fig. 2 are almost the same at various $V_{DC}$, and the transfer characteristic of the FET device fabricated with this tube exhibits metallic nature, corroborating the metallic behavior of its DFM response. In contrast, dramatic change of the DFM image for the nanotube shown in Fig. 3 is evident: the response is fairly weak at $V_{DC} = 3$ V, and grows gradually reaching a maximum at $V_{DC} = -3$ V, and then decreases monotonically as $V_{DC}$ is increased to 3 V. This transition is manifested quantitatively in Fig. 3(c). Intuitively, the DC bias plays the role of a local gate which tunes the majority carrier distribution in the tube and consequently modulates the strength of the DFM signal. The dielectric response increases (decreases) as the local carrier density in the segment below the AFM probe is raised (reduced). Our observation thus implies that the local carrier density increases as $V_{DC}$ is decreased, which is a signature of p-type behavior with holes as the majority carriers and is confirmed by the transfer characteristic of the corresponding FET device. In addition, more than twenty SWNTs have been studied with both DFM and FET and yielded similar results, solidly justifying our interpretation of the DFM response.

A Drude-level model is employed to understand the relation between the dielectric response and the microscopic properties of the material. The SWNT in DFM experiment is approximated as a one-dimensional conductor with carrier density $\rho$ and mobility $\mu$. When the AFM tip, which is represented with a point charge in the model, is place on top of the SWNT, the external field causes charge carriers to redistribute in the SWNT. The carrier (hole) density, $\rho(t, x)$, obeys the continuity equation

$$\frac{\partial \rho(t,x)}{\partial t} + \mu \frac{\partial}{\partial x}\left[\rho(t,x) E_x - \frac{k_B T_0}{e} \frac{\partial \rho(t,x)}{\partial x}\right] = 0 \qquad (1)$$

where $E_x$ is the effective electric field along the nanotube, $k_B$ is the Boltzman constant, $T_0$ is the temperature, and $e$ is the elementary charge. Figure 4(a) shows the carrier density profiles in the SWNT, at $t = 0$ and $3T/4$, where T is the period of the AC signal. The evolution of carrier density right beneath the tip from $t = 0$ to $t = T$ is shown in Fig. 4(a) inset.

The total force experienced by the AFM tip is:

$$F(t) = \frac{edC(V_{DC}+V_{AC}\sin(\omega t))}{4\pi\varepsilon_0} \int_{-L/2}^{L/2} \frac{\rho(t,x)}{(\sqrt{x^2+d^2})^3} dx \qquad (2)$$

where C is the capacitance of the system composed of the AFM tip and the nanotube, d is the height of the AFM tip relative to the nanotube, and $\varepsilon_0$ is the vacuum permittivity. The DFM force is obtained from the $2\omega$ component of the Fourier transformation of Eq. (2) (See Section 1 of the Electronic Supplementary Material for details). Figure 4(b) shows the numerically calculated dielectric force as a function of carrier density (black) and carrier mobility (red). The DFM force increases rapidly with the mean carrier density $\rho_0$ when $\rho_0 > 3 \times 10^5$ m$^{-1}$, indicating that a minimal carrier density is required to obtain a detectable dielectric response signal. The dependence of the dielectric force on the carrier mobility $\mu$ also agrees well with physical intuition: when $\mu$ is close to zero, the carriers are essentially bounded at their equilibrium position and thus do not contribute to the dielectric response; for $\mu > 10^{-8}$ cm$^2$/Vs, the DFM force rapidly increases with as $\mu$ increases, and saturates at $\mu \sim 1$ cm$^2$/Vs. This saturation is due to the finite length of the nanotube (See Section 2 of the Electronic Supplementary Material for details), which has also been experimentally observed previously [19]. As the dielectric force saturates at a fairly low mobility, the DFM signal is independent of $\mu$ for materials with high mobility such as SWNTs.

These results illustrate the intrinsic connection between transport characterization with FET devices and DFM measurements. The gate bias in both experiments modulates the charge carrier density in the SWNT; and the experimental observables, the source-drain current in the FET case and the dielectric response in the DFM case, are both critically dependent on carrier density and mobility. Such an underlying parallel forms the basis for the observed consistency between the DFM and the FET experiments on the same tubes. The major difference is that DFM is a contactless technique: the material under investigation is not electrically connected with an external circuit and thus there is no net carrier flow in or out of the material upon local gate modulation.

As both the dielectric response in DFM and the transfer characteristics in FET are determined by carrier density and mobility, they might be correlated through a simple relation. Such correlation is demonstrated in Fig. 5(a) for six semiconducting SWNTs: a semi-logarithmic relation is observed between the gate modulation ratio of the source-drain current $I_{max}/I_{min}$ in the FET device characterization and of the dielectric response $F_{max}/F_{min}$ in the DFM measurement. This correlation can be well understood employing the model described above and using the relation between the source-drain current density and the Fermi level of the SWNT [20], as detailed in Section 3 of the Electronic Supplementary Material.

This direct correlation between the gate modulation ratio in DFM and the current on/off ratio in FET has far-reaching implications: first of all, it indicates that the dielectric response in the DFM experiment is a true measurement of transfer characteristics, and parameters such as the gate modulation ratio reflect intrinsic material properties; secondly, from technological application point of view, the contactless DFM imaging technique is directly related to critical parameters in device applications, and thus can be used to predict the device performance before the material is fabricated into device.

It is worth noting that the parallel between DFM responses and transfer characteristics widely exists in many other aspects. We notice that transport properties of various metallic SWNT devices are essentially the same, as shown in Fig. 5(b), as long as fabrication recipe and contacts condition are unaltered. We thus choose metallic tubes as the "standard" and normalize the DFM response of semiconducting tubes with respect to that of a metallic tube measured with the same probe. The impact of tip-to-tip variation on the measured dielectric force is eliminated this way. In Fig. 5(c), DFM signals vs. $V_{DC}$ for three representative semiconducting SWNTs are shown, illustrating significant tube to tube variations. Comparing Fig. 5(c) to Fig. 5(d) which shows the corresponding conductance vs. gate voltage trace, we find that: (1) if the dielectric response of a SWNT increases monotonically (non-monotonically) with the decrease of $V_{DC}$, the device shows a unipolar (ambipolar) behavior; (2) the relative strength of the DFM signal among various tubes is consistent with the relative magnitude of transconductance of the corresponding devices; and (3) both DFM vs. $V_{DC}$ traces and transfer characteristics exhibit hysteresis of the same direction as a result of the charge trapping effect [21, 22]. These resemblances further certify that intrinsic carriers transport properties obtained from DFM are comprehensive and reliable.

As an SPM based technique possessing nanometer scale resolution, DFM is capable of detecting the location and nature of local transport barriers in nanotubes, which shed light on the complexity of transport behaviors in nanotube FETs. Tube A and Tube B in Fig. 6(a) are both laser ablation tubes with normal topography. The DFM response of Tube A in the depletion regime is homogeneous, however, a conspicuous semiconductor-semiconductor junction appears in the accumulation regime which might be a weak link that is less likely to survive a large current. Indeed, device fabricated with this tube breaks down at $I_d \sim 20$ nA. In contrast to Tube A, the DFM signal of Tube B is almost invariant along the tube in the accumulation regime, although multiple defects are observed in the depletion regime. Those barriers might not act as weak links for electronic transport, because with the most depletable section determining the current that could pass through the tube, the device is "off" with a negligibly small current in the depletion regime. As expected, the device made from this tube survives a current of $\sim 1$ µA. Tube C is a SWNT grown by CVD. All DFM images of the tube are uniform, and no noticeable defect is present. As a result, devices made from CVD tubes can survive a current up to 20 ~ 25 µA, reaching the limit of current carrying capability of SWNTs [1]. Fig. 6(b) shows the transfer characteristics of Tube B and Tube C. While the trace for Tube B is featured by a slow rise and several kinks, a sharp and smooth increase of drain current for Tube C is evident, confirming our interpretation of their distinguished DFM responses.

In summary, we have shown the underlying parallel between the FET transport and DFM measurements. The dielectric responses of SWNTs are highly correlated with their electronic transport properties. The relation between the dielectric force in the DFM experiment and the intrinsic carrier transport properties is revealed through a Drude-level model. For semiconducting tubes, a quantitative correlation between the DFM gate modulation ratio and FET device on/off ratio is established, which allows for device performance prediction based on contactless DFM imaging. Taking advantage of the spatial mapping capability of the DFM technique, we detected the location and the complex nature of defects in SWNTs and demonstrated their influences on the device transport behaviors. The explication of the physics behind the DFM measurement lay down the foundation for future application in characterizing the electronic properties of various nanomaterials.

Acknowledgments. This work was supported by the National Natural Science Foundation of China (Nos. 91233104 and 61376063) and the National Basic Research Program of China (2010CB934700). L. C. acknowledges the support from Jiangsu Provincial Natural Science Foundation (Grant No. BK20130006). The device fabrication and transport measurement were performed at the Nanofabrication Facility and Platform for Characterization and Test at Suzhou Institute of Nano-Tech and Nano-Bionics, Chinese Academy of Sciences. We thank Y. Wang, K. Hou, Y. Dong, B. Li, F. Tian, T. Zhou, and K. Huang for technical supports.

**Electronic Supplementary Material:** Supplementary material (Sections 1 & 2: details about the numerical model and calculation ; Section 3: the proof of the semi-logarithmic relation;  Section 4: the local variation of the DFM response in semiconducting SWNTs) is available in the online version of this article at http://dx.doi.org/10.1007/*************.

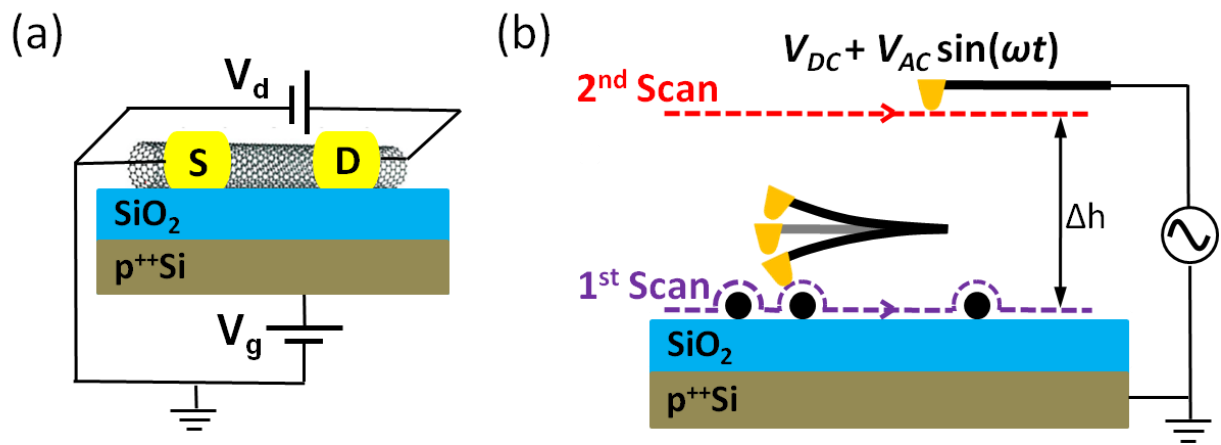

**Figure 1.** (a) Schematic illustration of the SWNT-FET device geometry and transport measurement. (b) Schematic illustration of the DFM experiment.

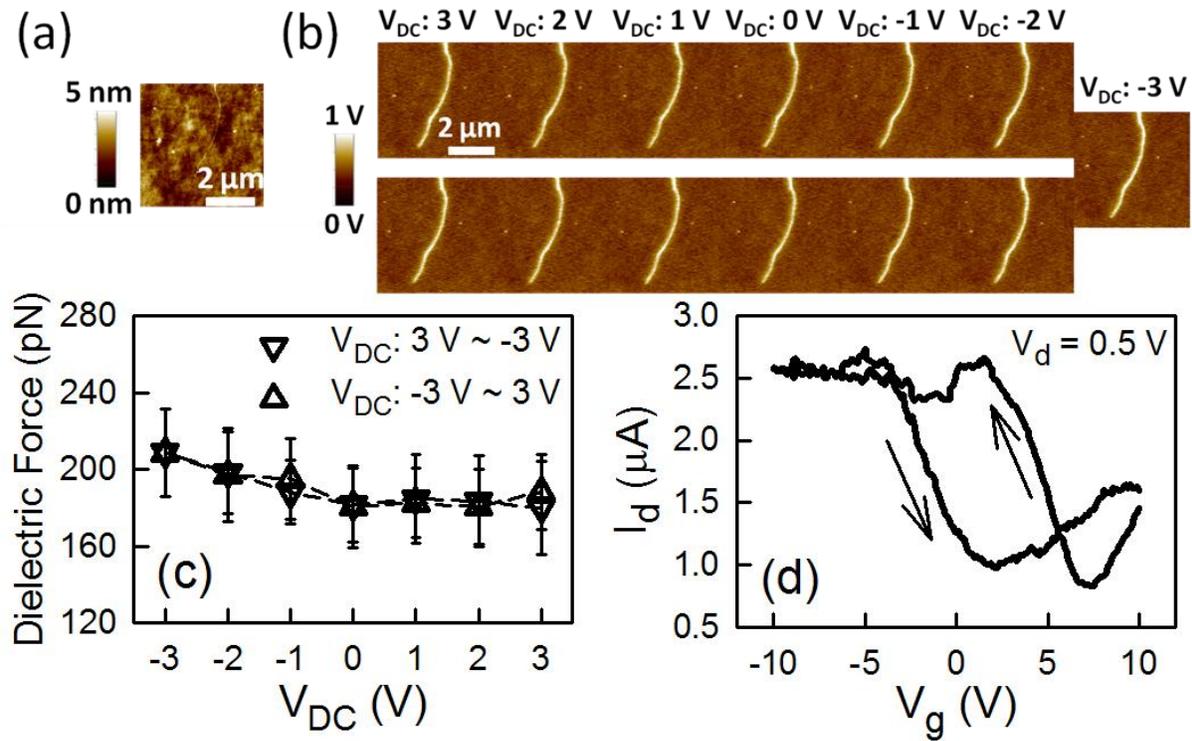

**Figure 2.** DFM and transport of a metallic nanotube. (a) Topographical AFM image of the tube. (b) Evolution of the DFM images of the tube in ambient air, as $V_{DC}$ changes from 3 V to -3 V and then back to 3 V. (c) Dielectric force as a function of $V_{DC}$. (d) Device transfer characteristic with a source-drain voltage $V_d = 0.5$ V.

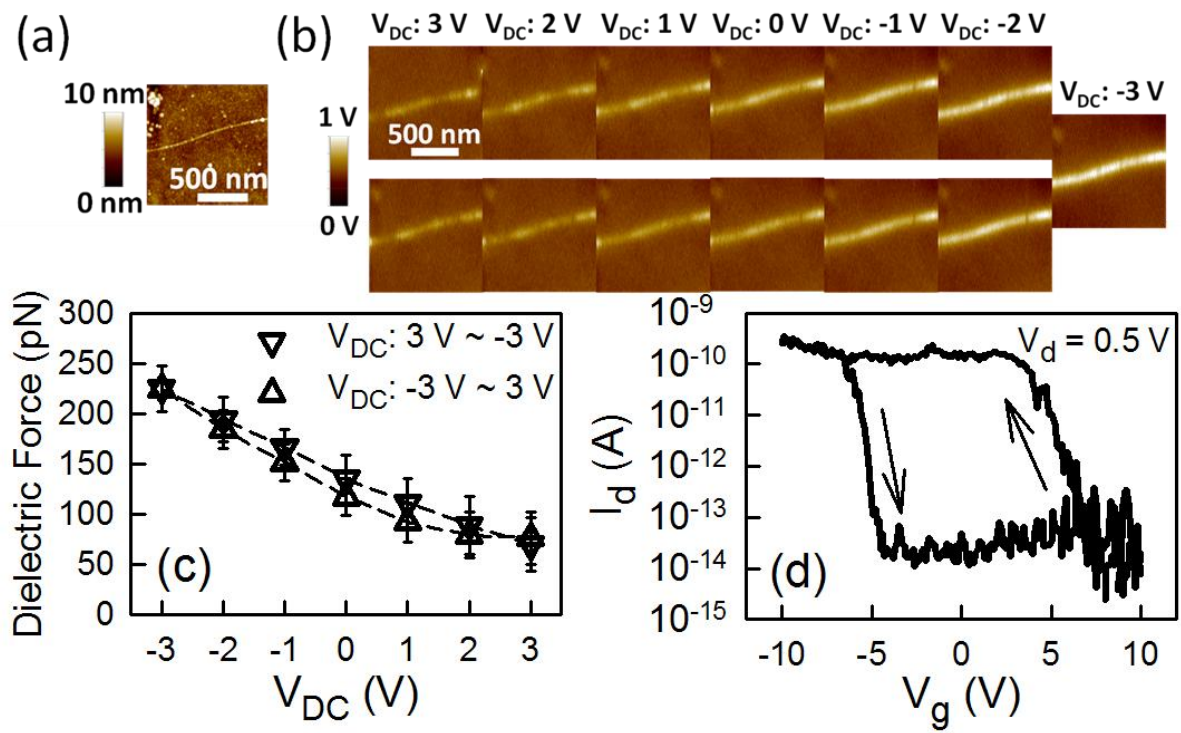

**Figure 3.** DFM and transport of a semiconducting nanotube. (a) Topographical AFM image of the tube. (b) Evolution of the DFM images of the tube in ambient air, as $V_{DC}$ changes from 3 V to -3 V and then back to 3 V. (c) Dielectric force as a function of $V_{DC}$. (d) Device transfer characteristic with a source-drain voltage $V_d = 0.5$ V.

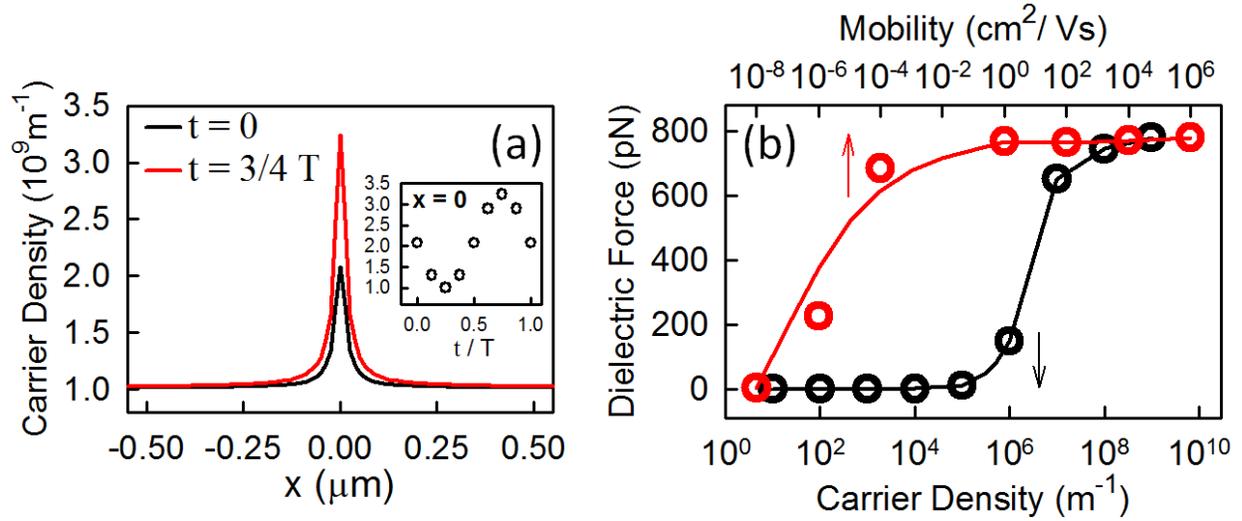

**Figure 4.** (a) Carrier density profiles at t = 0 (black) and t = 3/4 T (red). Inset shows the evolution of carrier density beneath the tip within one period of the AC signal. (b) The numerically calculated dielectric force as a function of carrier density (black) and mobility (red). Symbols are the numerical results and lines are guide to eyes. The parameters used to obtain the numerical results are L=10 μm, d=20 nm, C=5 aF, $V_{dc}$=-3.0 V, $V_{ac}$=3.0 V, f=10 kHz, ρ=$10^9$ $m^{-1}$(red), μ=$10^6$ $cm^2$/Vs (black).

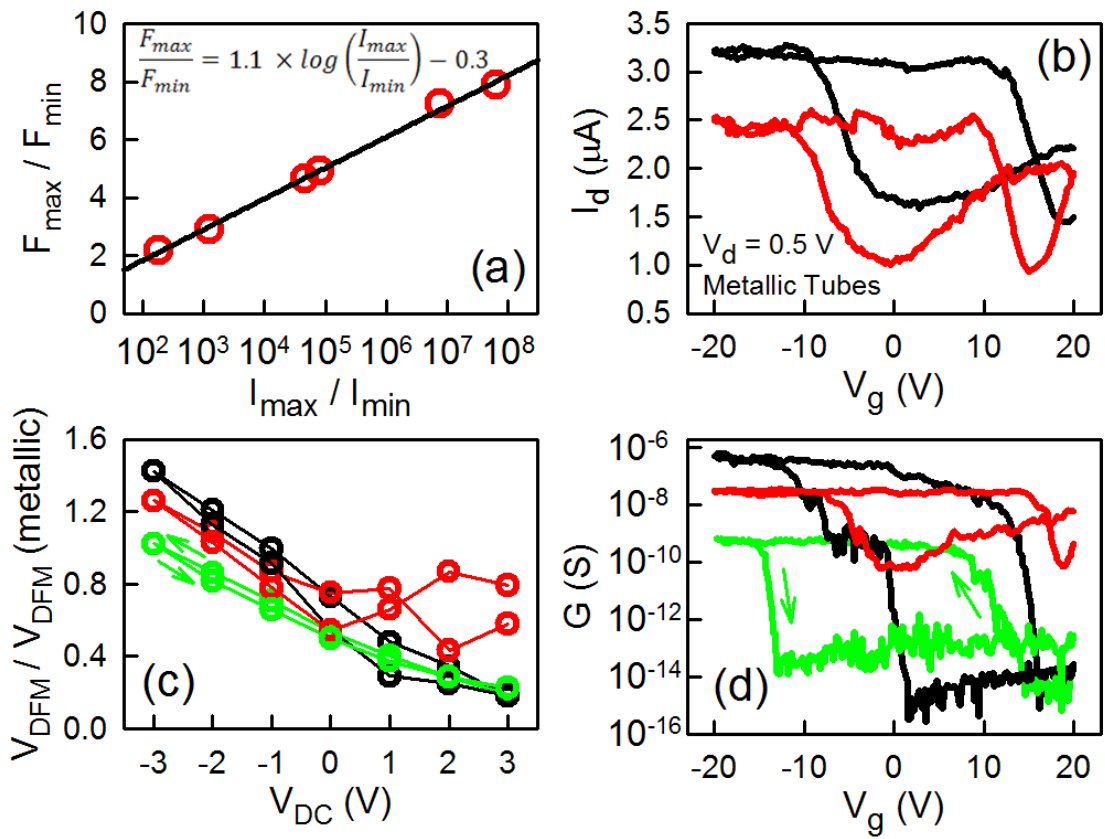

**Figure 5.** Correlations between the dielectric responses and transport properties for semiconducting SWNTs. (a) Dielectric force ratio of free tube as a function of the current ON/OFF ratio obtained from the transfer characteristics for six devices. Symbols are experimental data and solid line is the linear fit. (b) Transfer characteristics of two representative metallic tube devices at $V_d = 0.5$ V. (c) Normalized DFM response vs. $V_{DC}$ for three semiconducting tubes. (d) Conductance vs. gate voltage for devices fabricated with the three tubes shown in (c). The symbol colors in (c) and (d) are consistent.

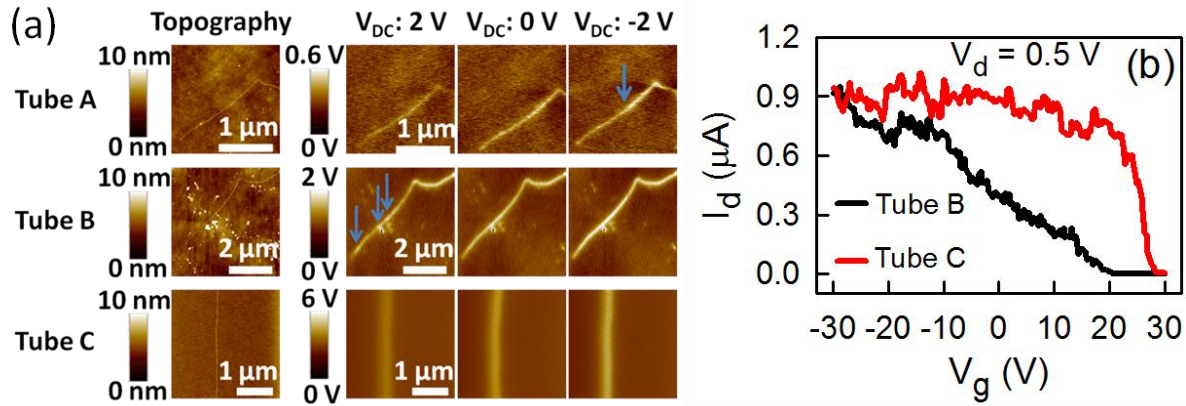

**Figure 6.** (a) Topography and DFM images of Tubes A, B and C. Tube A is a laser ablation tube with an intra S-S junction which is evident at $V_{DC}$ = -2 V. Tube B is another laser ablation tube: multiple defects are observed at $V_{DC}$ = 2 V (with a few representative ones pointed out by arrows). Tube C is a CVD tube where defects are hardly visible. (b) Transfer characteristics of Tube B and Tube C at $V_d$ = 0.5 V.

# Contactless Probing of the Intrinsic Carrier Transport in Single-Walled Carbon Nanotubes
# (Electronic Supplementary Material)

Yize Stephanie Li, Jun Ge, Jinhua Cai, Jie Zhang, Wei Lu, Jia Liu, and Liwei Chen

Suzhou Institute of Nano-Tech and Nano-Bionics, Chinese Academy of Sciences,

Suzhou, Jiangsu 215123, P. R. China

**Section 1: Numerical calculation of the dielectric force using the continuity equation**

Carrier density $\rho(t,x)$ obeys the continuity equation (Eq. (1)). The electric field $E$ in Eq. (1) is related to the charge density through Poisson equation

$$\nabla \cdot \mathbf{E} = \frac{e}{\varepsilon_0 \varepsilon_r} [\rho_{tip} + \rho(t,x) - \rho_{back}] \qquad (S1\text{-}1)$$

where $\rho_{tip}$ is the charge density on the AFM tip (a point charge distribution is assumed), and $\rho_{back}$ is the constant background charge density in the nanotube which makes the system charge neutral. The carrier density $\rho$ and the electric field $E$ are solved self-consistently from Eq. (1) and Eq. (S1-1).

Employing

$$\frac{\partial \rho(t,x)}{\partial t} \approx \frac{\rho(t,x) - \rho(t-\Delta t, x)}{\Delta t} \qquad (S1\text{-}2)$$

$$\frac{\partial \rho(t,x)}{\partial x} \approx \frac{\rho(t,x+\Delta x) - \rho(t,x-\Delta x)}{2\Delta x} \tag{S1-3}$$

$$\frac{\partial^2 \rho(t,x)}{\partial^2 x} \approx \frac{\rho(t,x+\Delta x) - 2\rho(t,x) + \rho(t,x-\Delta x)}{\Delta x^2} \tag{S1-4}$$

a set of discrete differential equations are established for numerical calculation. 200 uniformly portioned points for t in one period [0, T], and 800 uniformly portioned points for x in [-L/2, L/2] are used in our calculation.

Because the two ends of a nanotube are electrically isolated from any external circuit, the electric current J at both ends must be zero. Therefore, boundary conditions

$$\rho(t,x) = \rho(t+T,x) \tag{S1-5}$$

$$J(t, x = -L/2) = J(t, x = L/2) = 0 \tag{S1-6}$$

are employed to obtain the periodic solution of Eq. (1).

The force F(t) experienced by the AFM tip is calculated from Eq. (2). The 2ω components of F(t) are computed as

$$F_s(2\omega) = \frac{2}{T} \int_0^T F(t) \sin(2\omega t) dt \tag{S1-7}$$

$$F_c(2\omega) = \frac{2}{T} \int_0^T F(t) \cos(2\omega t) dt \tag{S1-8}$$

and the magnitude of 2ω component of F(t) is

$$F(2\omega) = \sqrt{F_s^2(2\omega) + F_c^2(2\omega)}. \tag{S1-9}$$

**Section 2: The saturation of the DFM force as a function of carrier mobility**

In the main text, we show that the dielectric force saturates when carrier mobility μ is large than 1 cm$^2$/Vs, and here we explain the reason. When μ is small, only the carriers in the vicinity of AFM tip take part in the motion driven by the applied electric field (Fig. S1(a)). As μ is increased, a larger number of carriers participate in the motion in response to the external field, and hence the dielectric force increases. However, once the carriers at the nanotube boundaries are also involved, no more extra carriers can respond to the applied electric field and contribute to the dielectric force, resulting in the saturation of the DFM force (Fig. S1(b)).

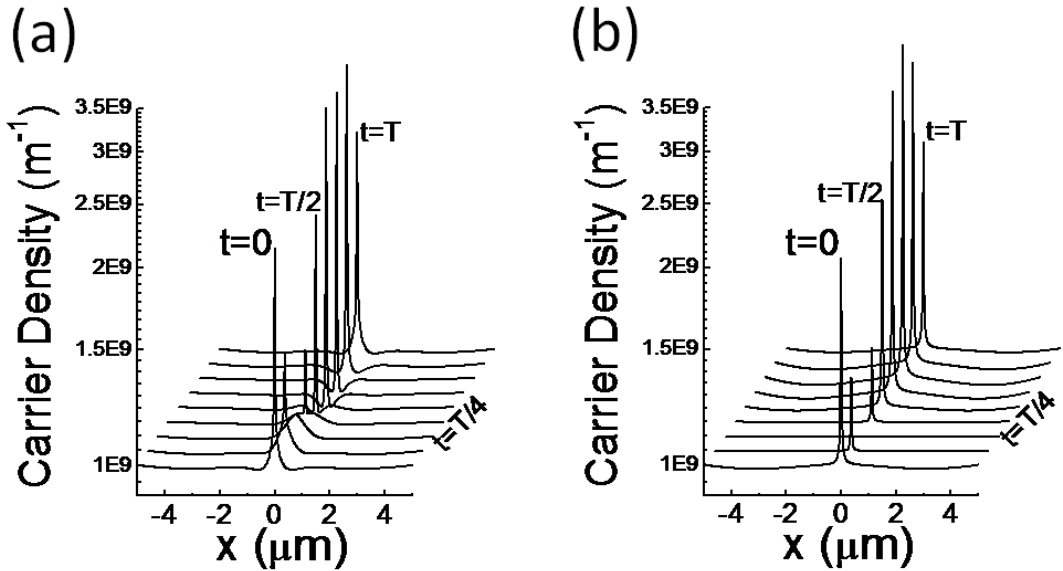

**Figure S1.** Evolution of the carrier density profile in one-dimensional material at different time within one period of the AC signal, for (a) μ = 10$^{-4}$ cm$^2$/Vs, and (b) μ = 1 cm$^2$/Vs. All the other parameters used in the numerical calculations are same as in Fig. 4. For the case shown in (b), the carriers accumulated around x = 0 at t = 0 are gradually released into the entire system in the 1st T/4 period, and a flat carrier distribution along the material is obtained at t = T/4, indicating that all carriers in the system are involved in the motion driven by the external electric field.

**Section 3: Proof of the semi-logarithmic relation revealed in Fig. 5(a):**

$$\frac{F_{max}}{F_{min}} \sim \log\left(\frac{I_{max}}{I_{min}}\right) + \text{const}$$

The source-drain current density of FET can be written as: $J_{SD}=\rho E\mu$, where E is the electrical field due to the source-drain voltage, and thus $\frac{I_{max}}{I_{min}} = \frac{\rho_{max} \cdot E \cdot \mu}{\rho_{min} \cdot E \cdot \mu} = \frac{\rho_{max}}{\rho_{min}}$. For a p-type semiconducting SWNT, the hole density $\rho = 2N_0 e^{\frac{E_v - E_F}{k_B \cdot T}}$, where $E_v$ is the valence band of SWNT and $N_0$ is the effective density of state [S1]. So:

$$\log\frac{I_{max}}{I_{min}} = \ln\frac{I_{max}}{I_{min}}/\ln 10 = -\frac{E_F(I_{max}) - E_F(I_{min})}{k_B \cdot T \cdot \ln 10} \tag{S3-1}$$

where $E_F(I_{max})$ and $E_F(I_{min})$ are the Fermi levels corresponding to $I_{max}$ and $I_{min}$ of the SWNT-FET, respectively.

From Eq. (1) and Eq. (2) in the main text and Section 1 in this Supplemental Material, the dielectric force can be written as

$$F \approx \frac{\rho_0 e^2 dC^2 V_{AC}^2}{2(4\pi\varepsilon_0)^2 \varepsilon_r k_B T} \int_{-L/2}^{L/2} \left(1 - \frac{3eC}{8\pi\varepsilon_0 \varepsilon_r k_B T} \cdot V_{DC} \cdot \frac{1}{\sqrt{x^2+d^2}}\right) \frac{1}{(x^2+d^2)^2} dx \tag{S3-2}$$

So:

$$\frac{F_{max}}{F_{min}} = 1 - \alpha (V_{DC,max} - V_{DC,min}) \tag{S3-3}$$

where $\alpha = \frac{3eC}{8\pi\varepsilon_0 \varepsilon_r k_B T} \frac{\int_{-L/2}^{L/2} \frac{dx}{\left(\sqrt{x^2+d^2}\right)^5}}{\int_{-L/2}^{L/2}\left(1-\frac{3eC}{8\pi\varepsilon_0\varepsilon_r k_B T}V_{DC,min}\frac{1}{\sqrt{x^2+d^2}}\right)\frac{dx}{(x^2+d^2)^2}}$ is a constant.

The total capacitance C is determined from the geometric capacitance $C_g$ and the quantum capacitance $C_q$ through the relation: $C^{-1} = C_g^{-1} + C_q^{-1}$, where $C = \frac{dQ}{dV_{DC}}$ and $C_q = \frac{dQ}{dE_F}$ [S2]. Therefore, the DC bias $V_{DC}$ and the Fermi level $E_F$ is correlated through [S1]:

$$E_F - E_{F0} = \frac{C_g}{C_g + C_q(E_F)} \cdot V_{DC} \tag{S3-4}$$

where $E_{F0}$ is the Fermi level of SWNT without DC bias. For low $V_{DC}$ value, which is the case for our experiment, $C_g$ is much larger than $C_q$, and thus Eq. (S3-4) is simplified to $E_F - E_{F0} = V_{DC}$. So Eq. (S3-3) can be rewritten as :

$$\frac{F_{max}(2\omega)}{F_{min}(2\omega)} \sim 1 - \alpha \left(E_{F_{max}} - E_{F_{min}}\right) \tag{S3-5}$$

where $E_{F_{max}}$ and $E_{F_{min}}$ are the Fermi level of the SWNT when DC bias is $V_{DC,max}$ and $V_{DC,min}$, respectively.

In a generic semiconductor, as the Fermi level shifts from the mid-gap energy level $E_{mid}$ to valence band edge $E_v$ or conductance band edge $E_c$, the charge carrier density increases. For p-type semiconducting SWNTs, the charge carrier density reaches maximum (minimum) when the Fermi level of SWNT is at $E_v$ ($E_{mid}$). As either the dielectric force or the source-drain current increases with the carrier density, we have $E_{F_{max}} \sim E_F(I_{max}) \sim E_v$ and $E_{F_{min}} \sim E_F(I_{min}) \sim E_{mid}$, and thus $\left(E_{F_{max}} - E_{F_{min}}\right) \sim \left(E_F(I_{max}) - E_F(I_{min})\right)$. Considering equations (S3-1) and (S3-5), the semi-logarithmic relation is then proven.

**Section 4: Local variation of the DFM response in semiconducting SWNTs**

As shown in Fig. S2(a) (which is the same as Fig. 3(b)), although the dielectric response in the accumulation regime (negative $V_{DC}$) does not vary much along the tube, local variation is apparent in the depletion regime (positive $V_{DC}$). In Fig. S2(b), the DFM gate modulation ratio for the three segments as indicated in Fig. S2(a) exhibits a significant variation. We suggest that it is the segment with the weakest dielectric signal that determines the carrier transport in the depletion regime, because it acts as a conduction bottleneck limiting the current carrying capability of the tube. Consequently, the macroscopic transport property is determined by the segment which has the largest DFM ratio, providing that the DFM response is uniform along the tube in the accumulation regime.

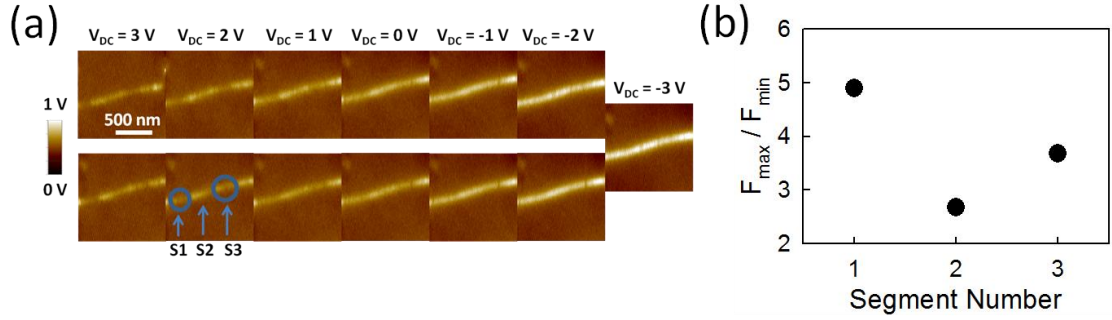

**Figure S2.** (a) Evolution of the DFM image of a semiconducting tube in ambient air, as $V_{DC}$ changes from 3 V to -3 V and then back to 3 V (same as Fig. 3(b), except that three segments with distinct DFM responses in the depletion regime are indicated). (b) Ratio of the dielectric forces at $V_{DC}$ = -3 V and 3 V for each of the three segments shown in (a). The dielectric force at $V_{DC}$ = 3 V is an average of the values obtained from downward and upward sweeps.

**Reference:**

[S1] J. Liang, D. Akinwande, H. -S. Philip Wong, J. Appl. Phys. **104**, 064515 (2008).

[S2] S. Ilani, L. A. K. Donev, M. Kindermann, and P. L. McEuen, Nat. Phys., **2**, 687 (2006).